\documentclass[twocolumn,prl]{revtex4}

\usepackage[T1]{fontenc}

\usepackage{graphicx}
\usepackage{amsmath,amssymb}
\usepackage{bm}
\usepackage{textcomp,gensymb,units}
\usepackage[usenames]{color}

\newcommand{\deriv}{\ensuremath{\mathrm{d}}}
\newcommand{\e}{\ensuremath{\mathrm{e}}}
\newcommand{\vc}[1]{\ensuremath{\bm{#1}}}

\newcommand{\ket}[1]{\ensuremath{|  #1 \rangle}}

\newcommand{\avg}[1]{\ensuremath{\langle {#1} \rangle}}
\newcommand{\ensembleavg}[1]{\ensuremath{\avg{#1}_e}}
\newcommand{\abs}[1]{\ensuremath{| {#1} |}}
\newcommand{\var}[1]{\ensuremath{\Delta{#1^2}}}
\newcommand{\Rb}{\ensuremath{{}^{87}\mathrm{Rb}}}

\newcommand{\Sy}{\ensuremath{S_y}}
\newcommand{\Sz}{\ensuremath{S_z}}
\newcommand{\Salpha}{\ensuremath{S_\alpha}}
\newcommand{\Smeanlength}{\ensuremath{\abs{\avg{\vc{S}}}}}
\newcommand{\Fone}{\ket{F=1,m_F=0}}
\newcommand{\Ftwo}{\ket{F=2,m_F=0}}
\newcommand{\Ntot}{\ensuremath{N_\text{tot}}}
\newcommand{\unitx}{\ensuremath{\vc{\hat x}}}
\newcommand{\unity}{\ensuremath{\vc{\hat y}}}
\newcommand{\unitz}{\ensuremath{\vc{\hat z}}}
\newcommand{\contrast}{\ensuremath{C}}
\newcommand{\snr}{\ensuremath{\sigma^2}}
\newcommand{\sqnoise}{\ensuremath{\snr_\text{min}}}
\newcommand{\asqnoise}{\ensuremath{\snr_\text{max}}}
\newcommand{\bothnoise}{\ensuremath{\snr_{\substack{\text{max} \\ \text{min}}}}}
\newcommand{\readoutnoise}{\ensuremath{\snr_\text{ro}}}

\newcommand{\sqclock}{\ensuremath{\zeta}}

\begin{document}

\title{Implementation of Cavity Squeezing of a Collective Atomic Spin}

\author{Ian D. Leroux}
\affiliation{
Department of Physics, MIT-Harvard Center for Ultracold Atoms
and Research Laboratory of Electronics, Massachusetts Institute of Technology,
Cambridge, Massachusetts 02139, USA}

\author{Monika H. Schleier-Smith}
\affiliation{
Department of Physics, MIT-Harvard Center for Ultracold Atoms
and Research Laboratory of Electronics, Massachusetts Institute of Technology,
Cambridge, Massachusetts 02139, USA}

\author{Vladan Vuleti\'{c}}
\affiliation{
Department of Physics, MIT-Harvard Center for Ultracold Atoms
and Research Laboratory of Electronics, Massachusetts Institute of Technology,
Cambridge, Massachusetts 02139, USA}

\date{\today}

\begin{abstract}
We squeeze unconditionally the collective spin of a dilute ensemble of
laser-cooled \Rb\ atoms using their interaction with a driven optical resonator.
The shape and size of the resulting spin uncertainty region are well described by a simple analytical model
[M.H.S., I.D.L., V.V., arXiv:0911.3936] through two orders of magnitude in the effective interaction strength, without free parameters.
We deterministically generate states with up to 5.6(6) dB of metrologically relevant spin squeezing on the canonical \Rb\ hyperfine clock transition.
\end{abstract}

\maketitle

Squeezed spin states~\cite{Kitagawa93,Wineland92,Wineland94,Sorensen01,Sorensen01BEC,Kuzmich98},
where a component of the total angular momentum of an ensemble of spins has less uncertainty~\cite{Kuzmich00,Takano09} than is
possible without quantum mechanical correlations~\cite{Meyer01,Esteve08,Schleier-Smith08,Appel09},
attract interest for both fundamental and practical reasons.
Fundamentally,
they allow the study of many-body entanglement but retain a simple description
in terms of a single collective angular-momentum variable~\cite{Sorensen01,Sorensen01BEC}.
Practically, they may be a means to overcome the projection noise limit on precision~\cite{Wineland92,Wineland94,Andre04,Santarelli99}.
Spin squeezing has been demonstrated using entanglement of ions via their shared motional modes~\cite{Meyer01},
repulsive interactions in a Bose-Einstein condensate~\cite{Esteve08},
or partial projection by measurement~\cite{Schleier-Smith08,Appel09}.

In a companion paper~\cite{Schleier-Smith09}
we propose a cavity feedback method for deterministic production of
squeezed spin states using light-mediated interactions between
distant atoms in an optical resonator.
This approach generates spin dynamics similar to those of the one-axis twisting Hamiltonian
$H\propto\Sz^2$
in Kitagawa and Ueda's original proposal~\cite{Kitagawa93}.
Cavity squeezing scales to much higher particle number than direct manipulation of ions~\cite{Meyer01}
(but see Ref.~\cite{Leibfried04} for a potentially scalable approach)
and employs dilute ensembles rather than dense condensates of interacting atoms~\cite{Esteve08}.
Unlike measurement-based squeezing~\cite{Schleier-Smith08,Appel09},
it unconditionally produces a known squeezed state independent of detector performance.

Here we implement cavity squeezing for the canonical
$\Fone\leftrightarrow\Ftwo$
hyperfine clock transition in \Rb\ atoms, achieving a
$\unit[5.6(6)]{dB}$
improvement in signal-to-noise ratio~\cite{Wineland92,Wineland94}.
To our knowledge, this is the largest such improvement to date.
Moreover, the shape and orientation of the uncertainty regions we observe agree with a straightforward analytical model~\cite{Schleier-Smith09}, without free parameters, over two orders of magnitude in effective interaction strength.

\begin{figure}
  \includegraphics[width=0.8\columnwidth]{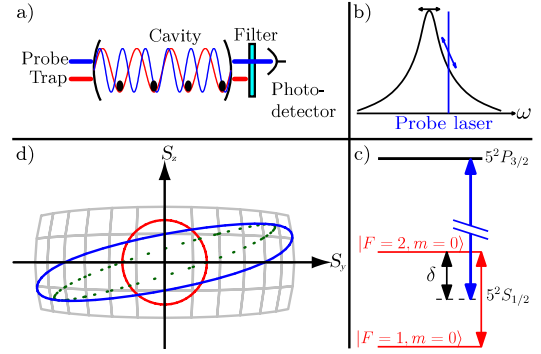}
  \caption{Cavity squeezing~\cite{Schleier-Smith09}:
    (a) The atoms are trapped in a standing-wave dipole trap inside an optical resonator.
    (b) The probe laser is detuned from cavity resonance by half a linewidth,
    so that atom-induced shifts of the cavity frequency change the transmitted power.
    (c) The cavity is tuned halfway between
    the optical transition frequencies for the two clock states.
    (d) The \Sz-dependent light shift shears the circular
    uncertainty region of the initial coherent spin state (red circle) into an
    ellipse (dotted).
    Photon shot noise causes phase broadening that increases
    the ellipse area (solid).
    The illustration is for a modest shearing $Q=3$ (see text). 
  }\label{fig:setup}
\end{figure}

Our scheme, similar in spirit to the proposal of Ref.~\cite{Takeuchi05},
relies on the repeated interaction of the atomic
ensemble with light circulating in an optical resonator, as
illustrated in Fig.~\ref{fig:setup}.
We label the two relevant eigenstates (clock states) of each one of
$N_0$ atoms as the spin-up and spin-down states of a spin-$1/2$
$\vc{s}_i$,
and define a total spin $\vc{S} = \sum_i \vc{s}_i$.
Its $z$ component corresponds to the population difference between clock states and its azimuthal
angle corresponds to their relative phase.
For a given total spin magnitude
$S=\abs{\vc{S}}\le S_0=N_0/2$
and a given permutation symmetry of the ensemble,
the set of possible collective states forms a Bloch sphere.

The coupling of the atoms to the resonator manifests itself both as a differential light shift
of the clock states which causes the $\vc{s}_i$ to precess about the \unitz\ axis,
and as a modified index of refraction which shifts the cavity resonance frequency. 
If a resonator mode is tuned halfway between the optical transition frequencies for
the two clock states [Fig.~\ref{fig:setup}(c)],
the atomic index of refraction produces opposite frequency shifts of the mode for atoms in each of the states,
yielding a net shift
$\Delta \omega_r / \kappa  = \phi_1 \Sz / 2 \ll 1$
proportional to the population difference $2\Sz$.
Here $\kappa$ is the linewidth of the resonator and
$\phi_1$ is the spin precession angle per photon transmitted through the resonator.
The resonator is driven by a probe laser with fixed incident power at a detuning
$\kappa/2$ 
so that this mode frequency shift changes the average number of photons transmitted by
$\Delta p = p_0 \phi_1 \Sz$ from its value $p_0$ in the absence of atoms.
As the intracavity power is \Sz-dependent,
so is the light shift, which produces a precession of each spin through an angle
$\phi(\Sz) = Q \Sz / S_0$.
The state of each atom now depends,
through \Sz,
on that of all other atoms in the ensemble.
The shearing strength
$Q = S_0 p_0 \phi_1^2$ is a dimensionless measure of the light-mediated interaction strength.
In particular, a coherent spin state prepared on the
equator of the Bloch sphere
(an uncorelated state with
$\avg{\vc{S}} = S\unitx$
and
$\var{\Sy}=\var{\Sz}=S/2$)
has its circular uncertainty region sheared into an ellipse with a shortened minor axis
[Fig.~\ref{fig:setup}(d)]~\cite{Schleier-Smith09}.

Two fundamental decoherence mechanisms counteract the unitary evolution which squeezes the spin uncertainty.
The first is photon shot noise:
the intracavity light field,
driven by a coherent input and decaying via the cavity mirrors,
is not in a photon number state and produces an uncertain light shift.
This uncertainty leads to irreversible phase broadening
$\var{\phi} = p_0 \phi_1^2 / 2 = Q / (2 S_0)$.
The squeezed variance,
which would lessen as
$Q^{-2}$
if the dynamics preserved the area of the uncertainty region,
therefore only decreases as
$Q^{-1}$~\cite{Schleier-Smith09}.

The second decoherence process is photon scattering into free space.
Scattered photons that reveal the state of individual atoms  spoil the ensemble's coherence,
while Raman scattering,
which changes the atoms' internal state at random,
increases the spin variance~\cite{Saffman09}.
In our system,
Rayleigh scattering occurs at the same rate for the two clock states,
does not reveal the atomic state,
and so does not harm the coherence~\cite{Ozeri05}.
At most 2.3\% of the atoms undergo Raman scattering for our parameters,
causing added noise and decoherence~\cite{Saffman09} much smaller than those from technical sources.

Finally,
the coherent shearing action ceases to reduce the minimum spin variance once the uncertainty region
becomes elongated enough that the curved geometry of the Bloch sphere becomes important~\cite{Kitagawa93}.
Such curvature effects are visible in our data for large values of the shearing strength
$Q$.

For the experimental demonstration, up to
$\Ntot=5\times10^{4}$
atoms of \Rb\
(with excited-state decay rate
$\Gamma=2\pi\times\unit[6.065]{MHz}$)
are confined in a standing-wave optical dipole trap inside a Fabry-P\'{e}rot resonator of linewidth
$\kappa=2\pi\times\unit[1.01(3)]{MHz}$.
Details of the apparatus are given in
Ref.~\cite{Schleier-Smith08}.
The atoms are coupled to the resonator with a position-dependent dimensionless cooperativity $\eta(\vc{r}) = 4 g(\vc{r})^2 / \kappa \Gamma$, where $2g(\vc{r})$ is the vacuum Rabi frequency~\cite{Kimble98}.
We define an effective cooperativity $\eta$ and atom number $N_0$ for an equivalent uniformly-coupled system so that the spin variance of a coherent spin state, measured via the resonator frequency shift, is $\var{\Sz}=S_0/2=N_0 / 4$.
The effective quantities must satisfy
$N_0\eta = \Ntot\ensembleavg{\eta(\vc{r})}$,
where \ensembleavg{} denotes an average over the ensemble, in order to reproduce the observed average frequency shift.
Reproducing the projection-noise-induced variance of the cavity shift requires
$N_0\eta^2 = \Ntot\ensembleavg{\eta(\vc{r})^2}$.
These two constraints impose the definitions
$\eta = \ensembleavg{\eta(\vc{r})^2} / \ensembleavg{\eta(\vc{r})} =0.139(5)$ and $N_0 = \Ntot \ensembleavg{\eta}/\eta$.
The effective total spin has
$\Sz = (N_2 - N_1)/2$,
where $N_{1,2}$ are the analogously-defined effective populations of
the clock states.

We prepare an initial coherent spin state
by optically pumping the atoms into \Fone\
($\avg{\vc{S}}=-S_0\unitz$) and applying a microwave $\pi/2$ pulse to yield
$\avg{\vc{S}}=S_0\unitx$.
The squeezing is
performed by two pulses of $\unit[780]{nm}$ light detuned
\unit[3.18(1)]{GHz} to the blue of the
$\ket{5^2S_{1/2},F=2}\leftrightarrow \ket{5^2P_{3/2},F'=3}$ transition
and \unit[500]{kHz} to the blue of the cavity resonance.
This yields a single-photon phase shift
$\phi_1=(2/3)\eta\Gamma / \delta=\unit[171(6)]{\micro rad}$,
where
$\delta=2\pi\times\unit[3.29]{GHz}$ is the effective detuning from the
$5^2P_{3/2}$
manifold with oscillator strength $2/3$.
Between the two optical pulses,
each of which lasts
$\unit[50]{\micro s}\gg\kappa^{-1}$
and contains up to
$\sim10^5$
photons,
is a composite
(SCROFULOUS~\cite{Cummins03})
microwave $\pi$ pulse, forming a spin echo sequence.
The spin echo cancels the spatially inhomogeneous phase shift caused by the
$p_0$
photons transmitted on average through the resonator but preserves the shearing effect.
We measure \Sz\ using this same sequence but with stronger optical pulses, with
$10^6$
photons transmitted on average.
The transmitted fraction of these pulses, measured on
an avalanche photodiode, reveals the cavity
resonance frequency shift and hence \Sz~\cite{Schleier-Smith08}.

\begin{figure}
  \includegraphics[width=0.8\columnwidth]{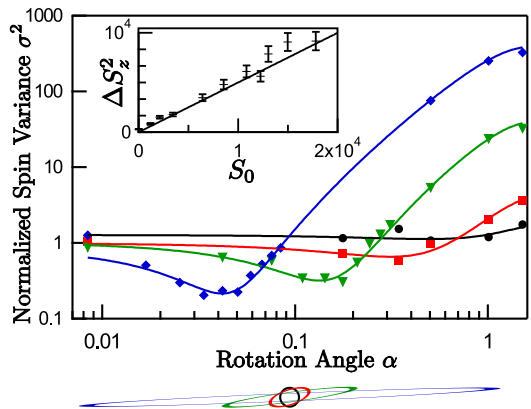}
  \caption{Normalized variance \snr\ as a function of 
    rotation angle $\alpha$ about the mean spin direction for states prepared with shearing
    $Q=0$ (black circles),
    $Q=1.2$ (red squares),
    $Q=7.7$ (green triangles)
    and $Q=30.7$ (blue diamonds).
    The curves are cosine fits.
    Statistical error bars are comparable to the symbol size.
    The shapes of the corresponding uncertainty regions are illustrated below the plot.
    Inset: Observed variance \var{\Sz}\ of the initial state as a function of
    $S_0$.
    The line is the projection noise limit as determined from cavity parameters.
  }\label{fig:ellipses}
\end{figure}

We observe the shearing by rotating the state through an angle
$\alpha$
with a microwave pulse about the axis of its mean spin vector and recording the variance
$\var{\Salpha}$
of a subsequent measurement of
$\Sz|_\alpha$
over a series of 100 identical preparations.
The measured variance is normalized to projection noise
$\snr(\alpha)=2\var{\Salpha} /S_0$
where $S_0$ is determined from our atom number calibration,
based on a first-principles calculation using accurately measured cavity parameters~\cite{Schleier-Smith08}.
The inset to Fig.~\ref{fig:ellipses} shows the observed
$\var{\Sz}$
of the initial spin state as a function of
$S_0$,
as well as the calculated projection noise limit. 

Typical data of \var{\Salpha}\ are displayed in Fig.~\ref{fig:ellipses}.
As the state is rotated the variance dips below projection noise as the minor axis of the
ellipse is aligned with \unitz\ and then increases beyond it as
the major axis in turn rotates towards \unitz.
The variation of $\var{\Salpha}$ with angle is sinusoidal with a period $\pi$, as it must be for any distribution of \Sy-\Sz\ fluctuations.
We record such curves over a range of photon numbers, corresponding to increasing shearing strength $Q$, keeping the effective atom number constant at $2S_0\approx 3.2\times10^{4}$.
We compare the fitted phase and minimum and maximum variance of each sinusoid to the predictions of our model~\cite{Schleier-Smith09}, briefly described below.

\begin{figure}
  \includegraphics[width=0.8\columnwidth]{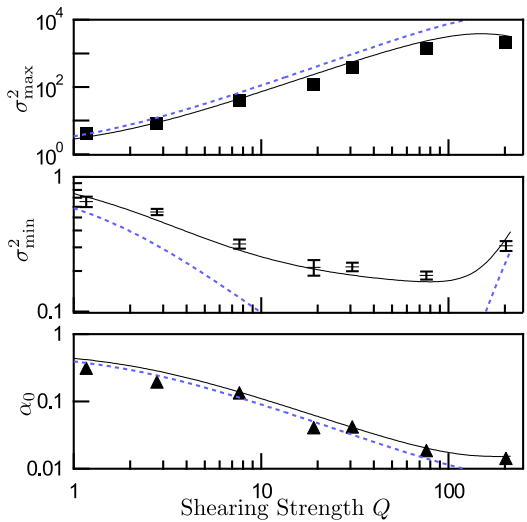}
  \caption{\asqnoise (top panel) is the normalized maximum variance,
    \sqnoise (middle panel) the normalized minimum variance, and
    $\alpha_0$ (bottom panel) the rotation angle for minimum variance
    for each measured ellipse as a function of shearing strength $Q$.
    Statistical error bars are given for \sqnoise, and are smaller
    than the symbols for \asqnoise\ and $\alpha_0$.  The blue dashed
    curves are theoretical predictions for an ideal system, while the
    black lines are predictions including separately measured
    technical imperfections.  Shearing was varied by adjusting the
    photon number $p_0 \approx 2200\times Q$.}\label{fig:modelling}
\end{figure}
Neglecting scattering,
the initial \Sz\ distribution with $\var{\Sz}=S_0/2$ is unaffected by the shearing,
while the \Sy\ variance is modified~\cite{Schleier-Smith09}
\begin{equation}
  \var{\Sy} = \frac{S^2}{2} + \frac{S_0}{4} -
                   \left(\frac{S^2}{2} - \frac{S_0}{4}\right)
                   \left(1 - \frac{\gamma Q}{S_0}\right)
                   \e^{-\frac{\xi^2 Q^2}{S_0}},
  \label{eqn:varsy}
\end{equation}
and an \Sy--\Sz\ correlation is introduced
\begin{equation}
  W
  = \avg{\Sy\Sz+\Sz\Sy}
  = \xi Q S \e^{-\frac{\xi^2 Q^2}{4 S_0}}.
  \label{eqn:sysz}
\end{equation}
The expressions given here are approximations to those of the
companion theory paper~\cite{Schleier-Smith09},
with which they agree to within 0.01\%
for our parameters.
They include additional correction factors
$\xi$, $\gamma$, and $S/S_0$ to account for technical imperfections.
$\xi= (\deriv\mathcal L / \deriv \omega_p) \kappa / (2\mathcal L)$
is the logarithmic derivative of the Lorentzian resonator transmission
$\mathcal L$ with respect to fractional probe detuning
$\omega_p/\kappa$.
Ideally,
$\omega_p = \kappa/2$ and
$\xi=1$ exactly.
As we do not maintain precisely this detuning,
$0.97\le\xi\le1$ for the data presented here.
The variance of the intracavity probe power in the absence of atoms, expressed as a multiple of photon shot noise, is
$\gamma = 1 + 2\var{p_f}/p_0 = 1 + p_0 / (8\times10^4) = 1 + Q / 37$,
where $\var{p_f}$ is the additional variance in the transmitted photon number caused by independently determined fractional light noise.
Finally,
dephasing reduces the effective radius of the Bloch sphere from $S_0$ to $S$,
measured from
the envelope amplitude of a Rabi nutation curve.
For modest shearing
$Q<20$
we maintain a Bloch sphere radius
$S > 0.80 S_0$,
but when $Q$ reaches $200$ the radius is reduced to
$S = 0.48 S_0$.

Equations~\ref{eqn:varsy} and \ref{eqn:sysz} together yield
a prediction for the observed normalized variance:
\begin{equation}
  \snr(\alpha)
  = \frac{1}{S_0}\left[V_+ - A\cos\left(2\alpha - 2\alpha_0\right)\right] + \readoutnoise
\end{equation}
where 
$V_\pm = \var{\Sy} \pm \var{\Sz}$,
$A = \sqrt{V_-^2+W^2}$,
and the rotation angle which minimizes the variance is
$\tan(2\alpha_0) = W/V_-$.
The additional variance of our imperfect readout
$\readoutnoise=0.13$
is determined by comparing successive measurements of the same state as in
Ref.~\cite{Schleier-Smith08}.

Figure~\ref{fig:modelling} shows the predictions of maximum and minimum variance
$\bothnoise=(V_+ \pm A)/S_0 + \readoutnoise$
and orientation $\alpha_0$ as black lines,
together with the data points extracted from the cosine fits.
It is remarkable that a simple analytical model,
without free parameters, provides good predictions of the shape,
area and orientation of the uncertainty region for values of the shearing strength $Q$ that span a factor of 200.
Note the increase of the minimum variance
\sqnoise\ for large shearing as the curvature of the Bloch sphere becomes important.
For comparison,
Fig.~\ref{fig:modelling} also includes the model predictions without technical corrections
($S=S_0$, $\gamma=\xi=1$, $\readoutnoise=0$)
as blue dashed curves.
The maximal noise \asqnoise\ and the ellipse angle $\alpha_0$,
dominated by the shearing-induced broadening along \unity,
are insensitive to technical effects,
but the minimum variance \sqnoise\ is strongly affected by readout noise \readoutnoise.
This noise is due to a combination of finite quantum efficiency and avalanche noise of the photodetector together with Raman scattering which limits the number of photons in the readout measurement.
It could be suppressed by using a different photodetector to remove the avalanche noise and by performing the readout near-detuned to a cycling transition to suppress Raman scattering~\cite{Saffman09}.

\begin{figure}
  \includegraphics[width=0.8\columnwidth]{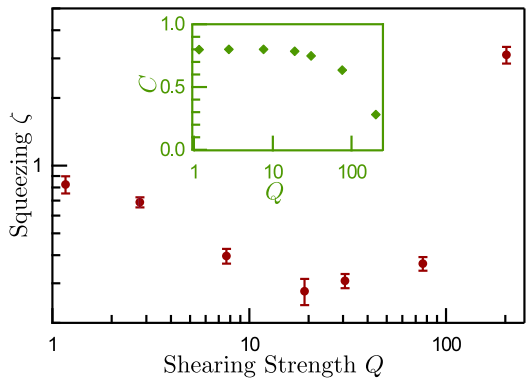}
  \caption{Metrologically-relevant squeezing \sqclock\ (red solid circles)
    and signal contrast \contrast\ (green diamonds, inset) as a function of
    shearing strength $Q$.
  }\label{fig:squeezing}
\end{figure}

To determine whether the reduced spin noise \sqnoise\ allows a gain in spectroscopic precision,
we must establish a signal-to-noise ratio by comparing it to the mean spin signal \Smeanlength~\cite{Wineland92}.
\Smeanlength\ is reduced below
$S_0$
by dephasing,
which shortens the Bloch vector,
and by shot-to-shot phase fluctuations which reduce the average projection of
$\vc{S}$
along its mean direction.
We specify the signal strength by a contrast
$\contrast=\abs{\avg{\vc{S}}}/S_0$,
measured from the mean amplitude of Rabi oscillations with the sheared state as input,
and plotted in the inset to Fig.~\ref{fig:squeezing}.
The signal contrast in the absence of squeezing light
$\contrast_\text{in}=0.80$ is limited by dephasing from the lock light used to stabilize the cavity length~\cite{Schleier-Smith08}.

Figure~\ref{fig:squeezing} shows the metrological squeezing parameter~\cite{Wineland92}
$\sqclock = 2\Smeanlength_\text{in} / (\Smeanlength^2 / \var{\Sz})
= \sqnoise \contrast_\text{in} / \contrast^2$, which
compares the squared signal-to-noise ratio for an ideal projection-noise-limited measurement using the initial spin signal
$\Smeanlength_\text{in} = \contrast_\text{in} S_0$
to that for the minimum observed variance and corresponding signal
$\Smeanlength$.
$\sqclock < 1$ indicates an improvement in signal-to-noise ratio
unattainable without entanglement~\cite{Wineland92}.
For $S_0 = 1.6\times10^{4}$ and $p_0=4.1\times10^4$
($Q=19$),
we reduce the spin noise by a factor
$\sqnoise=\unit[-6.7(6)]{dB}$.
At this photon number our contrast is
$\contrast = 0.78(2)$,
so that we demonstrate a
$\sqclock^{-1}=\unit[5.6(6)]{dB}$
improvement in potential measurement precision over that of the initial uncorrelated state.

By subtracting our independently measured readout noise \readoutnoise\ from \sqnoise,
we infer that states prepared by a shearing
$Q=19$ have an intrinsic spin variance that is a full
$\unit[10(1)]{dB}$
below the projection noise limit.
However,
it is the observed and not the intrinsic variance that determines the precision of a spectroscopic measurement,
and we use the former in calculating the spin noise reduction and the squeezing
$\sqclock$.

In conclusion,
we have demonstrated a method for deterministically generating squeezed states using switchable light-mediated interactions
in a dilute ensemble of otherwise non-interacting atoms.
Our model predicts the size and shape of the uncertainty region when technical effects are included.
We hope to observe states with substantially lower spin noise by improving our readout,
and to demonstrate their use in an atomic clock.

This work was supported in part by the NSF, DARPA,
and the NSF Center for Ultracold Atoms.
M.~H.~S. acknowledges support from the Hertz Foundation and NSF.
I.~D.~L. acknowledges support from NSERC.


\begin{thebibliography}{21}
\expandafter\ifx\csname natexlab\endcsname\relax\def\natexlab#1{#1}\fi
\expandafter\ifx\csname bibnamefont\endcsname\relax
  \def\bibnamefont#1{#1}\fi
\expandafter\ifx\csname bibfnamefont\endcsname\relax
  \def\bibfnamefont#1{#1}\fi
\expandafter\ifx\csname citenamefont\endcsname\relax
  \def\citenamefont#1{#1}\fi
\expandafter\ifx\csname url\endcsname\relax
  \def\url#1{\texttt{#1}}\fi
\expandafter\ifx\csname urlprefix\endcsname\relax\def\urlprefix{URL }\fi
\providecommand{\bibinfo}[2]{#2}
\providecommand{\eprint}[2][]{\url{#2}}

\bibitem[{\citenamefont{Kitagawa and Ueda}(1993)}]{Kitagawa93}
\bibinfo{author}{\bibfnamefont{M.}~\bibnamefont{Kitagawa}} \bibnamefont{and}
  \bibinfo{author}{\bibfnamefont{M.}~\bibnamefont{Ueda}},
  \bibinfo{journal}{Phys.\ Rev.\ A} \textbf{\bibinfo{volume}{47}},
  \bibinfo{pages}{5138} (\bibinfo{year}{1993}).

\bibitem[{\citenamefont{Wineland et~al.}(1992)\citenamefont{Wineland,
  Bollinger, Itano, Moore, and Heinzen}}]{Wineland92}
\bibinfo{author}{\bibfnamefont{D.~J.} \bibnamefont{Wineland}},
  \bibinfo{author}{\bibfnamefont{J.~J.} \bibnamefont{Bollinger}},
  \bibinfo{author}{\bibfnamefont{W.~M.} \bibnamefont{Itano}},
  \bibinfo{author}{\bibfnamefont{F.~L.} \bibnamefont{Moore}}, \bibnamefont{and}
  \bibinfo{author}{\bibfnamefont{D.~J.} \bibnamefont{Heinzen}},
  \bibinfo{journal}{Phys.\ Rev.\ A} \textbf{\bibinfo{volume}{46}},
  \bibinfo{pages}{R6797} (\bibinfo{year}{1992}).

\bibitem[{\citenamefont{Wineland et~al.}(1994)\citenamefont{Wineland,
  Bollinger, Itano, and Heinzen}}]{Wineland94}
\bibinfo{author}{\bibfnamefont{D.~J.} \bibnamefont{Wineland}},
  \bibinfo{author}{\bibfnamefont{J.~J.} \bibnamefont{Bollinger}},
  \bibinfo{author}{\bibfnamefont{W.~M.} \bibnamefont{Itano}}, \bibnamefont{and}
  \bibinfo{author}{\bibfnamefont{D.~J.} \bibnamefont{Heinzen}},
  \bibinfo{journal}{Phys.\ Rev.\ A} \textbf{\bibinfo{volume}{50}},
  \bibinfo{pages}{R67} (\bibinfo{year}{1994}).

\bibitem[{\citenamefont{S\o{}rensen and M\o{}lmer}(2001)}]{Sorensen01}
\bibinfo{author}{\bibfnamefont{A.~S.} \bibnamefont{S\o{}rensen}}
  \bibnamefont{and}
  \bibinfo{author}{\bibfnamefont{K.}~\bibnamefont{M\o{}lmer}},
  \bibinfo{journal}{Phys. Rev. Lett.} \textbf{\bibinfo{volume}{86}},
  \bibinfo{pages}{4431} (\bibinfo{year}{2001}).

\bibitem[{\citenamefont{S{\o}rensen et~al.}(2001)\citenamefont{S{\o}rensen,
  Duan, Cirac, and Zoller}}]{Sorensen01BEC}
\bibinfo{author}{\bibfnamefont{A.}~\bibnamefont{S{\o}rensen}},
  \bibinfo{author}{\bibfnamefont{L.-M.} \bibnamefont{Duan}},
  \bibinfo{author}{\bibfnamefont{J.~I.} \bibnamefont{Cirac}}, \bibnamefont{and}
  \bibinfo{author}{\bibfnamefont{P.}~\bibnamefont{Zoller}},
  \bibinfo{journal}{Nature} \textbf{\bibinfo{volume}{409}}, \bibinfo{pages}{63}
  (\bibinfo{year}{2001}).

\bibitem[{\citenamefont{Kuzmich et~al.}(1998)\citenamefont{Kuzmich, Bigelow,
  and Mandel}}]{Kuzmich98}
\bibinfo{author}{\bibfnamefont{A.}~\bibnamefont{Kuzmich}},
  \bibinfo{author}{\bibfnamefont{N.~P.} \bibnamefont{Bigelow}},
  \bibnamefont{and} \bibinfo{author}{\bibfnamefont{L.}~\bibnamefont{Mandel}},
  \bibinfo{journal}{Europhys. Lett.} \textbf{\bibinfo{volume}{42}},
  \bibinfo{pages}{481} (\bibinfo{year}{1998}).

\bibitem[{\citenamefont{Kuzmich et~al.}(2000)\citenamefont{Kuzmich, Mandel, and
  Bigelow}}]{Kuzmich00}
\bibinfo{author}{\bibfnamefont{A.}~\bibnamefont{Kuzmich}},
  \bibinfo{author}{\bibfnamefont{L.}~\bibnamefont{Mandel}}, \bibnamefont{and}
  \bibinfo{author}{\bibfnamefont{N.~P.} \bibnamefont{Bigelow}},
  \bibinfo{journal}{Phys.\ Rev.\ Lett.} \textbf{\bibinfo{volume}{85}},
  \bibinfo{pages}{1594} (\bibinfo{year}{2000}).

\bibitem[{\citenamefont{Takano et~al.}(2009)\citenamefont{Takano, Fuyama,
  Namiki, and Takahashi}}]{Takano09}
\bibinfo{author}{\bibfnamefont{T.}~\bibnamefont{Takano}},
  \bibinfo{author}{\bibfnamefont{M.}~\bibnamefont{Fuyama}},
  \bibinfo{author}{\bibfnamefont{R.}~\bibnamefont{Namiki}}, \bibnamefont{and}
  \bibinfo{author}{\bibfnamefont{Y.}~\bibnamefont{Takahashi}},
  \bibinfo{journal}{Phys. Rev. Lett.} \textbf{\bibinfo{volume}{102}},
  \bibinfo{pages}{033601} (\bibinfo{year}{2009}).

\bibitem[{\citenamefont{Meyer et~al.}(2001)\citenamefont{Meyer, Rowe,
  Kielpinski, Sackett, Itano, Monroe, and Wineland}}]{Meyer01}
\bibinfo{author}{\bibfnamefont{V.}~\bibnamefont{Meyer}},
  \bibinfo{author}{\bibfnamefont{M.~A.} \bibnamefont{Rowe}},
  \bibinfo{author}{\bibfnamefont{D.}~\bibnamefont{Kielpinski}},
  \bibinfo{author}{\bibfnamefont{C.~A.} \bibnamefont{Sackett}},
  \bibinfo{author}{\bibfnamefont{W.~M.} \bibnamefont{Itano}},
  \bibinfo{author}{\bibfnamefont{C.}~\bibnamefont{Monroe}}, \bibnamefont{and}
  \bibinfo{author}{\bibfnamefont{D.~J.} \bibnamefont{Wineland}},
  \bibinfo{journal}{Phys. Rev. Lett.} \textbf{\bibinfo{volume}{86}},
  \bibinfo{pages}{5870} (\bibinfo{year}{2001}).

\bibitem[{\citenamefont{Est\'{e}ve et~al.}(2008)\citenamefont{Est\'{e}ve,
  Gross, Weller, Giovanazzi, and Oberthaler}}]{Esteve08}
\bibinfo{author}{\bibfnamefont{J.}~\bibnamefont{Est\'{e}ve}},
  \bibinfo{author}{\bibfnamefont{C.}~\bibnamefont{Gross}},
  \bibinfo{author}{\bibfnamefont{A.}~\bibnamefont{Weller}},
  \bibinfo{author}{\bibfnamefont{S.}~\bibnamefont{Giovanazzi}},
  \bibnamefont{and} \bibinfo{author}{\bibfnamefont{M.~K.}
  \bibnamefont{Oberthaler}}, \bibinfo{journal}{Nature}
  \textbf{\bibinfo{volume}{455}}, \bibinfo{pages}{1216} (\bibinfo{year}{2008}).

\bibitem[{\citenamefont{Schleier-Smith
  et~al.}(2008)\citenamefont{Schleier-Smith, Leroux, and
  Vuleti\'{c}}}]{Schleier-Smith08}
\bibinfo{author}{\bibfnamefont{M.~H.} \bibnamefont{Schleier-Smith}},
  \bibinfo{author}{\bibfnamefont{I.~D.} \bibnamefont{Leroux}},
  \bibnamefont{and}
  \bibinfo{author}{\bibfnamefont{V.}~\bibnamefont{Vuleti\'{c}}},
  \emph{\bibinfo{title}{Reduced-quantum-uncertainty states for an atomic
  clock}} (\bibinfo{year}{2008}),
  \bibinfo{note}{arXiv:0810.2582, to be published in Phys. Rev. Lett.}.

\bibitem[{\citenamefont{Appel et~al.}(2009)\citenamefont{Appel, Windpassinger,
  Oblak, Hoff, Kj{\ae}rgaard, and Polzik}}]{Appel09}
\bibinfo{author}{\bibfnamefont{J.}~\bibnamefont{Appel}},
  \bibinfo{author}{\bibfnamefont{P.~J.} \bibnamefont{Windpassinger}},
  \bibinfo{author}{\bibfnamefont{D.}~\bibnamefont{Oblak}},
  \bibinfo{author}{\bibfnamefont{U.~B.} \bibnamefont{Hoff}},
  \bibinfo{author}{\bibfnamefont{N.}~\bibnamefont{Kj{\ae}rgaard}},
  \bibnamefont{and} \bibinfo{author}{\bibfnamefont{E.~S.}
  \bibnamefont{Polzik}}, \bibinfo{journal}{Proceedings of the National Academy
  of Sciences} \textbf{\bibinfo{volume}{106}}, \bibinfo{pages}{10960}
  (\bibinfo{year}{2009}).

\bibitem[{\citenamefont{Andr\'e et~al.}(2004)\citenamefont{Andr\'e,
  S\o{}rensen, and Lukin}}]{Andre04}
\bibinfo{author}{\bibfnamefont{A.}~\bibnamefont{Andr\'e}},
  \bibinfo{author}{\bibfnamefont{A.~S.} \bibnamefont{S\o{}rensen}},
  \bibnamefont{and} \bibinfo{author}{\bibfnamefont{M.~D.} \bibnamefont{Lukin}},
  \bibinfo{journal}{Phys. Rev. Lett.} \textbf{\bibinfo{volume}{92}},
  \bibinfo{pages}{230801} (\bibinfo{year}{2004}).

\bibitem[{\citenamefont{Santarelli et~al.}(1999)\citenamefont{Santarelli,
  Laurent, Lemonde, Clairon, Mann, Chang, Luiten, and Salomon}}]{Santarelli99}
\bibinfo{author}{\bibfnamefont{G.}~\bibnamefont{Santarelli}},
  \bibinfo{author}{\bibfnamefont{P.}~\bibnamefont{Laurent}},
  \bibinfo{author}{\bibfnamefont{P.}~\bibnamefont{Lemonde}},
  \bibinfo{author}{\bibfnamefont{A.}~\bibnamefont{Clairon}},
  \bibinfo{author}{\bibfnamefont{A.~G.} \bibnamefont{Mann}},
  \bibinfo{author}{\bibfnamefont{S.}~\bibnamefont{Chang}},
  \bibinfo{author}{\bibfnamefont{A.~N.} \bibnamefont{Luiten}},
  \bibnamefont{and} \bibinfo{author}{\bibfnamefont{C.}~\bibnamefont{Salomon}},
  \bibinfo{journal}{Phys.\ Rev.\ Lett.} \textbf{\bibinfo{volume}{82}},
  \bibinfo{pages}{4619} (\bibinfo{year}{1999}).

\bibitem[{\citenamefont{Schleier-Smith
  et~al.}(2009)\citenamefont{Schleier-Smith, Leroux, and
  Vuleti\'{c}}}]{Schleier-Smith09}
\bibinfo{author}{\bibfnamefont{M.~H.} \bibnamefont{Schleier-Smith}},
  \bibinfo{author}{\bibfnamefont{I.~D.} \bibnamefont{Leroux}},
  \bibnamefont{and}
  \bibinfo{author}{\bibfnamefont{V.}~\bibnamefont{Vuleti\'{c}}},
  \emph{\bibinfo{title}{Squeezing the collective spin of a dilute atomic
  ensemble by cavity feedback}} (\bibinfo{year}{2009}),
  \bibinfo{note}{arXiv:0911.3936, to be published in  Phys. Rev. A}.

\bibitem[{\citenamefont{Leibfried et~al.}(2004)\citenamefont{Leibfried,
  Barrett, Schaetz, Britton, Chiaverini, Itano, Jost, Langer, and
  Wineland}}]{Leibfried04}
\bibinfo{author}{\bibfnamefont{D.}~\bibnamefont{Leibfried}},
  \bibinfo{author}{\bibfnamefont{M.~D.} \bibnamefont{Barrett}},
  \bibinfo{author}{\bibfnamefont{T.}~\bibnamefont{Schaetz}},
  \bibinfo{author}{\bibfnamefont{J.}~\bibnamefont{Britton}},
  \bibinfo{author}{\bibfnamefont{J.}~\bibnamefont{Chiaverini}},
  \bibinfo{author}{\bibfnamefont{W.~M.} \bibnamefont{Itano}},
  \bibinfo{author}{\bibfnamefont{J.~D.} \bibnamefont{Jost}},
  \bibinfo{author}{\bibfnamefont{C.}~\bibnamefont{Langer}}, \bibnamefont{and}
  \bibinfo{author}{\bibfnamefont{D.~J.} \bibnamefont{Wineland}},
  \bibinfo{journal}{Science} \textbf{\bibinfo{volume}{304}},
  \bibinfo{pages}{1476} (\bibinfo{year}{2004}).

\bibitem[{\citenamefont{Takeuchi et~al.}(2005)\citenamefont{Takeuchi, Ichihara,
  Takano, Kumakura, Yabuzaki, and Takahashi}}]{Takeuchi05}
\bibinfo{author}{\bibfnamefont{M.}~\bibnamefont{Takeuchi}},
  \bibinfo{author}{\bibfnamefont{S.}~\bibnamefont{Ichihara}},
  \bibinfo{author}{\bibfnamefont{T.}~\bibnamefont{Takano}},
  \bibinfo{author}{\bibfnamefont{M.}~\bibnamefont{Kumakura}},
  \bibinfo{author}{\bibfnamefont{T.}~\bibnamefont{Yabuzaki}}, \bibnamefont{and}
  \bibinfo{author}{\bibfnamefont{Y.}~\bibnamefont{Takahashi}},
  \bibinfo{journal}{Phys. Rev. Lett.} \textbf{\bibinfo{volume}{94}},
  \bibinfo{eid}{023003} (\bibinfo{year}{2005}).

\bibitem[{\citenamefont{Saffman et~al.}(2009)\citenamefont{Saffman, Oblak,
  Appel, and Polzik}}]{Saffman09}
\bibinfo{author}{\bibfnamefont{M.}~\bibnamefont{Saffman}},
  \bibinfo{author}{\bibfnamefont{D.}~\bibnamefont{Oblak}},
  \bibinfo{author}{\bibfnamefont{J.}~\bibnamefont{Appel}}, \bibnamefont{and}
  \bibinfo{author}{\bibfnamefont{E.~S.} \bibnamefont{Polzik}},
  \bibinfo{journal}{Phys. Rev. A} \textbf{\bibinfo{volume}{79}},
  \bibinfo{pages}{023831} (\bibinfo{year}{2009}).

\bibitem[{\citenamefont{Ozeri et~al.}(2005)\citenamefont{Ozeri, Langer, Jost,
  DeMarco, Ben-Kish, Blakestad, Britton, Chiaverini, Itano, Hume
  et~al.}}]{Ozeri05}
\bibinfo{author}{\bibfnamefont{R.}~\bibnamefont{Ozeri}},
  \bibinfo{author}{\bibfnamefont{C.}~\bibnamefont{Langer}},
  \bibinfo{author}{\bibfnamefont{J.~D.} \bibnamefont{Jost}},
  \bibinfo{author}{\bibfnamefont{B.}~\bibnamefont{DeMarco}},
  \bibinfo{author}{\bibfnamefont{A.}~\bibnamefont{Ben-Kish}},
  \bibinfo{author}{\bibfnamefont{B.~R.} \bibnamefont{Blakestad}},
  \bibinfo{author}{\bibfnamefont{J.}~\bibnamefont{Britton}},
  \bibinfo{author}{\bibfnamefont{J.}~\bibnamefont{Chiaverini}},
  \bibinfo{author}{\bibfnamefont{W.~M.} \bibnamefont{Itano}},
  \bibinfo{author}{\bibfnamefont{D.~B.} \bibnamefont{Hume}},
  \bibnamefont{et~al.}, \bibinfo{journal}{Phys. Rev. Lett.}
  \textbf{\bibinfo{volume}{95}}, \bibinfo{eid}{030403} (\bibinfo{year}{2005}).

\bibitem[{\citenamefont{Kimble}(1998)}]{Kimble98}
\bibinfo{author}{\bibfnamefont{H.~J.} \bibnamefont{Kimble}},
  \bibinfo{journal}{Physica Scripta} \textbf{\bibinfo{volume}{T76}},
  \bibinfo{pages}{127} (\bibinfo{year}{1998}).

\bibitem[{\citenamefont{Cummins et~al.}(2003)\citenamefont{Cummins, Llewellyn,
  and Jones}}]{Cummins03}
\bibinfo{author}{\bibfnamefont{H.~K.} \bibnamefont{Cummins}},
  \bibinfo{author}{\bibfnamefont{G.}~\bibnamefont{Llewellyn}},
  \bibnamefont{and} \bibinfo{author}{\bibfnamefont{J.~A.} \bibnamefont{Jones}},
  \bibinfo{journal}{Phys. Rev. A} \textbf{\bibinfo{volume}{67}},
  \bibinfo{pages}{042308} (\bibinfo{year}{2003}).

\end{thebibliography}

\end{document}